# Giant orbital diamagnetism of three-dimensional Dirac electrons in Sr$_3$PbO antiperovskite

S. Suetsugu[1,2,3], K. Kitagawa[1], T. Kariyado[4], A. W. Rost[2,5,6], J. Nuss[2], C. Mühle[2], M. Ogata[1] and H. Takagi[1,2,5]

[1]Department of Physics, The University of Tokyo, 7-3-1 Hongo, Bunkyo-ku, Tokyo 113-0033, Japan

[2]Max Planck Institute for Solid State Research, Heisenbergstrasse 1, 70569 Stuttgart, Germany

[3]Department of Physics, Kyoto University, Kyoto 606-8502, Japan.

[4]International Center for Materials Nanoarchitectonics, National Institute for Materials Science, Tsukuba 305-0044, Japan

[5]Institute for Functional Matter and Quantum Technologies, University of Stuttgart, Pfaffenwaldring 57, 70550 Stuttgart, Germany

[6]SUPA, School of Physics and Astronomy, University of St Andrews, North Haugh, St Andrews, Fife KY16 9SS, United Kingdom

## Abstract

In Dirac semimetals, inter-band mixing has been known theoretically to give rise to a giant orbital diamagnetism when the Fermi level is close to the Dirac point. In Bi$_{1-x}$Sb$_x$ and other Dirac semimetals, an enhanced diamagnetism in the magnetic susceptibility $\chi$ has been observed and interpreted as a manifestation of such giant orbital diamagnetism. Experimentally proving their orbital origin, however, has remained challenging. Cubic antiperovskite Sr$_3$PbO is a three-dimensional Dirac electron system and shows the giant diamagnetism in $\chi$ as in the other Dirac semimetals. $^{207}$Pb NMR measurements are conducted in this study to explore the microscopic origin of diamagnetism. From the analysis of the Knight shift $K$ as a function of $\chi$ and the relaxation rate $T_1^{-1}$ for samples with different hole densities, the spin and the orbital components in $K$ are successfully separated. The results establish that the enhanced diamagnetism in Sr$_3$PbO originates from the orbital contribution of Dirac electrons, which is fully consistent with the theory of giant orbital diamagnetism.

Dirac semimetals (*1, 2*), whose band crossing is protected by the crystalline symmetry, have attracted considerable interest, largely because of the expected topological properties. The formation of unusual surface states is a direct consequence of the nontrivial topology of their band structure with Dirac dispersions (*3*). The Berry curvature around the Dirac node gives rise to unconventional responses to magnetic fields such as a nontrivial phase shift in quantum oscillations (*4, 5*) and a chiral anomaly (*6, 7*). So far, the research effort has focused mainly on such intra-band Berry curvature and related physics. However, it may be tempting to note that inter-band effects should not be dismissed here; a nontrivial topology of band-mixing between the conduction and the valence bands can lead to an inter-band Berry connection (*8*) and gives rise to exotic phenomena as in the case for the intra-band topological effect. The giant orbital diamagnetism of Dirac electrons (*9, 10*) may be viewed as such an inter-band topological effect. This mechanism is distinct from the other kinds of magnetisms originating from itinerant electrons, Pauli paramagnetism and Landau diamagnetism, which are scaled by the density of states of itinerant electrons at the Fermi level.

A large diamagnetism of the order of $10^{-4}$ emu/mol, larger than the expected Larmor core diamagnetism of $\sim 10^{-5}$ emu/mol, was first recognized long ago in the measurements of bulk magnetic susceptibility in semimetal Bi (*11*), which is known to host massive three-dimensional Dirac bands with a small number of electrons. The large diamagnetism of Bi was found to be further enhanced by Sb-doping in $Bi_{1-x}Sb_x$ (*12, 13*), which reduces the number of electrons in the Dirac bands (*14*). Eventually the diamagnetism is maximized when the Fermi level $E_F$ lies in the Dirac mass-gap (Fig. 1A) with Sb $x>x_c=0.07$. Recently discovered three-dimensional (3D) Dirac semimetals also show a large diamagnetism of similar magnitude in the bulk magnetic susceptibility (*15–18*), when their $E_F$ is located near the Dirac points.

The microscopic origin of the giant diamagnetism in Dirac semimetals, in particular, $Bi_{1-x}Sb_x$, has been a subject of intense theoretical debates over decades. The early attempts to understand it as a Landau diamagnetism failed to explain the maximized diamagnetic response when $E_F$ lies in the gap. The orbital magnetism of Dirac bands in the presence of inter-band effects was then proposed to be the origin (*9*), which can explain the enhancement of diamagnetism towards the Dirac point reasonably and has been accepted as the theoretical picture behind the giant diamagnetism. One of the intuitive pictures for the theoretical understanding of giant orbital diamagnetism is based on the $E$-linear density of states $D(E_\perp)$, where $E_\perp$ is a two-dimensional kinetic energy for the momentum perpendicular to the applied field. When Dirac electrons are confined in Landau levels under a magnetic field, the average energy gain and loss are not balanced, in contrast to the case for ordinary parabolic bands with constant $D(E_\perp)$ (Fig. 1B) (*10*), which increases the total free energy of Dirac bands under a magnetic field and gives rise to a diamagnetism. Note that the orbital diamagnetism comprises the contributions from all the electrons occupying the Dirac bands, not only from the electrons around the $E_F$, as in the Pauli spin paramagnetism and the Landau (orbital) diamagnetism. This picture naturally explains the maximally enhanced orbital susceptibility when the $E_F$ lies in the Dirac mass gap.

Despite the progress in the theoretical understanding of the origin of giant diamagnetism, its experimental verification has remained challenging, as it requires a separation of the orbital component from the spin component. The expected orbital diamagnetism from Dirac electrons of the order of $10^{-4}$ emu/mol is large but still could be comparable to the spin Pauli paramagnetism, for example, when bands other than the Dirac bands contribute and/or the *g*-factor is enhanced from 2. Magnetic resonance techniques in principle could analyze the contributions from different origins. The microscopic magnetism of $Bi_{1-x}Sb_x$ has been studied by nuclear magnetic resonance (NMR) (*19*), muon spin rotation (μSR) (*20*, *21*), perturbed angular distribution (*22*, *23*) and β-NMR (*24*). The verification of the orbital character of diamagnetism using these techniques, however, has been far from complete. In the case of NMR, the large electric quadrupole interaction of a nuclear spin $I \geq 1$ in $^{209}$Bi NMR has imposed critical constraints on the detailed analysis of the electronic contribution and the separation of spin and orbital contributions. NMR study on an $I = 1/2$ nuclear spin system, without electric quadrupole and phonon interactions, should be a promising approach to verify the orbital origin of the giant diamagnetism in Dirac semimetals. Dirac semimetals containing appropriate nuclear species, however, have been limited.

$Sr_3PbO$, the material we study here, is a member of the antiperovskite family $A_3Tt$O ($A$=Ca, Sr, Ba, Eu; $Tt$=Si, Ge, Sn, Pb) (*25*) and is theoretically proposed to be a three-dimensional massive Dirac electron system (*26*, *27*) with topological surface states (*28*, *29*). The cubic antiperovskite structure of Sr$_3$PbO is shown in Fig. 2A, where the Pb atoms are on the corners of cubic unit cell and the Sr atoms form an octahedron surrounding the O atom at the center. In the ionic limit, the valence states of constituent ions can be expressed as Sr$^{2+}_3$Pb$^{4-}$O$^{2-}$. In the reported band structures (*26*), the valence and the conduction bands indeed consist of the fully occupied 6$p$ orbitals of Pb$^{4-}$ and the empty 4$d$ orbitals of Sr$^{2+}$, respectively. The 6$p$ and the 4$d$ bands overlap marginally and a gap opens almost everywhere on the band crossing plane. The $C_4$ rotational symmetry, however, protects the band crossing at six equivalent points on Γ-X lines (Fig. 2B), which leads to six moderately anisotropic 3D Dirac bands free from the other parabolic bands. The 3D Dirac band has a very small mass gap of ~10 meV, which is created by the admixture of higher energy orbital states via spin-orbit coupling. The six Dirac bands merge at –125 meV below the Dirac points, giving rise to a saddle point. Below the saddle point, multi-band Fermi surfaces are expected to emerge when the Fermi level lies in this region. This region is essentially away from the Dirac physics. The presence of 3D Dirac electrons in Sr$_3$PbO is supported by recent experiments (*18*, *30*) which show the presence of extremely light mass (~ 0.01$m_e$) holes and $B$-linear magnetoresistance. Angle-resolved photoemission spectroscopy on a sister compound Ca$_3$PbO confirms the Dirac dispersion of the valence band predicted by band calculations (*31*).

The Pb antiperovskite should provide a promising arena for NMR studies of Dirac semimetals to verify the orbital character of the giant diamagnetism from Dirac electrons, as $^{207}$Pb hosts $I$ = 1/2 nuclear moment in contrast to $^{209}$Bi. Here, we report $^{207}$Pb NMR and magnetic susceptibility $\chi$ studies of the 3D Dirac system Sr$_3$PbO. An enhanced diamagnetism is observed in the magnetic susceptibility $\chi$ as in Bi and other 3D Dirac systems. Using the Korringa relation with the spin-lattice relaxation rate $T_1^{-1}$, we show that the spin contribution $K_{spin}$ in $K$ cannot account for the enhanced diamagnetism. The $K$-$\chi$ plot can be analyzed as the superposition of the spin and the orbital contributions with distinct hyperfine coupling constants, consistently with the analysis of the Korringa relation. The estimated orbital hyperfine constant indicates the delocalized nature of electrons in charge of the large orbital susceptibility. These results strongly affirm that the enhanced diamagnetism originates from the giant orbital susceptibility of Dirac electrons.

Five polycrystalline samples of $Sr_3PbO$ from different batches A-E with different hole densities ($p$) from ~ $10^{18}$ to ~ $10^{20}$ cm$^{-3}$ were investigated. The Hall resistivity $\rho_{xy}$ in the zero-field limit gives positive Hall coefficients $R_H$ = +3.8, 0.13, 0.032 and 0.029 and cm$^{-3}$/C (Fig. 2C), yielding hole densities $p$ = 1.6×10$^{18}$, 5.0×10$^{19}$, 2.0×10$^{20}$ and 2.2×10$^{20}$ cm$^{-3}$ for samples A, C, D and E, respectively. Sample B should have a comparable but only slightly smaller $p$ than sample C, judging from the NMR data. The donors very likely correspond to 0.01-1% level of cation defects and/or excess oxygens, which are introduced partially to relax the extremely reduced anionic state of $Pb^{4-}$. The result of band calculations in Fig. 2D indicates that the experimentally observed hole densities correspond to $E_F$ being –45, –125, –235 and –250 meV (measured from the center of the mass gap) for samples A, C, D and E respectively. These $E_F$s are indicated by the dashed lines in the schematic band picture shown in Fig. 2E. The $E_F$ of sample A lies in the Dirac band region while that of sample C (and B) is around the saddle point and those of samples D and E are in the multi Fermi surface region below the saddle point. The Dirac physics should manifest itself almost exclusively in sample A.

The magnetic susceptibilities $\chi(T)$ for three samples A, C and E, shown in Fig. 2F, are found to be all diamagnetic. The magnitude of diamagnetism increases with decreasing the hole concentration $p$ and hence increasing $E_F$ from E to A. The increase from samples C and E to sample A with $E_F$ in the Dirac bands is particularly significant and as large as of the order of 10$^{-4}$ emu/mol, which is comparable to the large diamagnetism observed in other Dirac semimetals (*15, 16*). An appreciable temperature dependence is observed particularly for sample A. The $\chi(T)$ of sample A shows a clear decrease with lowering temperature to ~30 K, which should be an intrinsic behavior of magnetic susceptibility. This is followed by a Curie-like increase very likely associated with magnetic impurities (0.01% level of $s$ = 1/2 impurities) at low temperatures. The other samples with a high hole density ($E_F$) show a monotonic increase of $\chi(T)$ from room temperature down to 2 K, with a clear Curie-like behavior of similar magnitude as sample A at low temperatures. It is not possible, however, to fit the $\chi(T)$ behavior for samples C and E over entire temperature range only with a Curie-Weiss contribution and a constant offset, particularly at high temperatures above 100 K. This indicates the presence of very weak but appreciable increase of the intrinsic $\chi(T)$ (broken lines in Fig. 2F) with lowering $T$ in samples C and E at least at high temperatures above 100 K where the Curie contribution is negligibly small. Note that the weak temperature dependence of the intrinsic $\chi(T)$ for samples C and E is negative, opposite to that of sample A.

The enhanced diamagnetism in sample A with the $E_F$ in the Dirac bands should represent the same large diamagnetism observed in $Bi_{1-x}Sb_x$ and other Dirac semimetals, which cannot be described naively by the conventional kinds of magnetisms. The core diamagnetism is estimated to be $-8.5\times10^{-5}$ emu/mol for $Sr_3PbO$ (32), which should not depend appreciably on the 1% level of cation defects or excess oxygens. The Pauli paramagnetism calculated from the density of states in the band calculation is only of the order of $10^{-5}$ emu/mol assuming $p \sim 2\times10^{20}$ cm$^{-3}$ and $g = 2$ for sample E, not as large as the difference of susceptibilities between samples A and E. At this point, however, the possibility of an enhanced $g$-factor, which could account for the difference, cannot be excluded completely.

$^{207}$Pb NMR measurements for samples A-E were conducted to verify the orbital origin of giant diamagnetism experimentally. The NMR spectra at 150 K are shown in Fig. 3A. A systematic shift of the peak position as a function of $p$ (and hence $E_F$) is observed from sample A to E, implying that the NMR peaks originate from the bulk $Sr_3PbO$. We note that the observed shifts are different from those of possible impurity phases such as 1.081%, –0.034% and 0.444% for metallic Pb, PbO and $PbO_2$ (33), respectively. The presence of sub-peaks in samples A and B are attributed to inhomogeneity/phase separation where small region(s) with a slightly different hole density from the main phase coexists (see also fig. S3).

The $p$- and $T$-dependence of the NMR Knight shift $K(T)$ (Fig. 3B), determined from the NMR spectra, is intimately related to those of the bulk magnetic susceptibility $\chi(T)$ shown in Fig. 2F. $K(T)$ decreases with decreasing $p$ from sample E to sample A, particularly from sample B and C to sample A. With increasing temperature, $K(T)$ for sample A shows an appreciable increase while samples E-B show a very weak decrease of $K(T)$ in parallel to those of $\chi(T)$. The close correlation between $K(T)$ and $\chi(T)$ indicates that $K(T)$ captures the $p$- and $T$-dependence of $\chi(T)$ including the enhanced diamagnetism. The separation of spin and orbital contributions in $K(T)$ should provide a clue to identify the orbital origin of enhanced diamagnetism.

The NMR Knight shift $K$ is comprised of several contributions as $K(T) = K_{\text{chem}} + K_{\text{spin}}(T) + K_{\text{orb}}(T)$, essentially the same as the bulk magnetic susceptibility. $K_{\text{spin}}$ and $K_{\text{orb}}$ are spin and orbital contributions, respectively. Each term is proportional to the respective susceptibilities $\chi_{spin}$ and $\chi_{orb}$ via the respective hyperfine coupling constants $A_{spin}$ and $A_{orb}$ as $K_{spin} = A_{spin}\chi_{spin}/N_A\mu_B$ and $K_{orb} = A_{orb}\chi_{orb}/N_A\mu_B$. Here, $N_A$ and $\mu_B$ are the Avogadro constant and the Bohr magneton, respectively (hereinafter we omit $N_A\mu_B$). $A_{spin}$ and $A_{orb}$ are in principle different. The chemical shift $K_{chem}$ does not depend on $T$ and $p$ and gives a constant offset to $K$. The relationship between $K(T)$ and $\chi(T)$ for samples A, C and E, the $K$-$\chi$ plot, is shown in Fig. 4A, confirming the close correlation between $K(T)$ and $\chi(T)$. To exclude the influence from the extrinsic Curie contribution, we plot here data only for $T > 50$K. The $K(T)$-$\chi(T)$ relationship for each sample is linear with almost common slopes among different samples. These straight lines for different samples, however, shift upward upon going from samples A to E with decreasing $E_F$ and do not fall onto a universal $K(T)$-$\chi(T)$ line, which strongly suggests that both $\chi_{spin}$ and $\chi_{orb}$ give appreciable contributions to the observed $K(T)$ with distinct $A_{spin}$ and $A_{orb}$. The $T$-dependence of $K(T)$ is highly likely dominated by one of the two contributions, as the $K(T)$-$\chi(T)$ slope originating from the $T$-dependence is universal. A correlation between $K(T)$ and $\chi(T)$ that is always positive implies that both $A_{spin}$ and $A_{orb}$ are positive.

The Korringa behavior of the spin-lattice relaxation rate $T_1^{-1}$, $(T_1T)^{-1}$ = constant, is observed for all samples A-E over a wide temperature range as shown in Fig. 3C, which provides us with important hints to estimate $K_{spin}$ and hence $K_{orb}$. The magnitude of $(T_1T)^{-1}$, which is proportional to the square of the electron density of states $D(E_F)$ at $E_F$, increases systematically with increasing $p$ (decreasing $E_F$) from samples A to E, which can be reasonably understood as the increase of $D(E_F)$. Indeed, as seen in Fig. 3D, $(T_1T)^{-1/2}$ at 100 K for the samples with different $E_F$ are scaled well with the calculated $D(E_F)$ for the 6$p$ orbitals of Pb, implying that $(T_1T)^{-1/2}$ captures the $D(E_F)$ that determines the spin contribution in $\chi(T)$ and $K(T)$. For sample A with the lowest hole density, a clear upward deviation from the Korringa behavior can be seen at high temperatures, which can be reasonably ascribed to the $T$-dependence of the thermally averaged density of states around $E_F$, $\langle D(E)^2 \rangle_T$ (34–36). Assuming strongly $E$-dependent $D(E) \propto (E-E_{DP})^2$ around the Dirac points at $E = E_{DP}$, the $T$-dependence of $T_1^{-1}$ for sample A can be reproduced well (solid line in Fig. 3C), yielding $E_F$–$E_{DP}$ ~ –60 meV (open square in Fig. 3D), roughly consistent with that estimated from the hole concentration and the band calculation (for details, see (37)). We also note that the orbital contribution to $T_1^{-1}$ was theoretically estimated to be at least an order of magnitude smaller than the observed spin contribution (38), which can be neglected here.

Confirming that $(T_1T)^{-1/2}$ is a good measure of $D(E_F)$, we can estimate roughly the spin contribution $K_{spin}$ in the observed $K$. The Korringa relation, $T_1TK_{spin}^2 = S$, yields the linear dependence of $K_{spin}$ on $(T_1T)^{-1/2}$ with the slope $S^{1/2}$. For a simple metal with an isotropic Fermi surface, the Korringa value $S$ is determined by the gyromagnetic-ratio of the nuclei under observation, $\gamma_n$, and the gyromagnetic-ratio of an electron, $\gamma_e$, as $S = \hbar/4\pi k_B \times (\gamma_e/\gamma_n)^2$. The $p$-dependence of the $g$-factor could modify the Korringa relationship as both $K_{spin}$ and $T_1^{-1}$ are in proportion to the square of $g$ (*37*). The excellent scaling of $T_1^{-1}$ with $D(E_F)$ over a wide variety of $p$ values, however, indicates that the $p$-dependence of the $g$-factor is not appreciable within the range of $p$ investigated here and that $g$ is unlikely to be strongly modified from 2. The plot of $K$ as a function of $(T_1T)^{-1/2}$ is shown in Fig. 4B. $K$ decreases with decreasing $(T_1T)^{-1/2}$, a measure of the density of states, from sample E to sample A, much more rapidly than the expected linear relationship $K_{spin} = S^{1/2}(T_1T)^{-1/2}$ (gray dashed line). The strongly non-linear decrease of $K$ from sample E can be naturally explained by the superposition of an additional orbital contribution $K_{orb}$, which increases rapidly with increasing $E_F$ (decreasing $p$) towards the Dirac mass gap.

It is known that the effective Korringa value $S^*$, inferred from an experimentally observed slope of $K$-$(T_1T)^{-1/2}$, is often larger than $S$ calculated from the gyromagnetic-ratio by more than a factor of 2 even in a simple metal. The black dashed line with an enhanced $S^* \sim 6.3S$ in Fig. 4B connects the data for all the heavily doped samples from E to B. This assumes that $K$ is fully dominated by $K_{spin}$ ($K_{orb} \sim 0$) for these samples and therefore gives the upper-bound estimate for $K_{spin}$. Considering that the $E_F$s of samples E-B are outside the Dirac band regime, the assumption of $K_{orb} \sim 0$ for them highly likely captures the reality better than the $S^* = S$ limit. Even with the maximum estimate of $K_{spin}$, Fig. 4B clearly indicates that a large orbital contribution must be incorporated to account for the enhanced diamagnetic $K$ for sample A.

Taking a closer look at Fig. 4B, we recognize the two important features to identify the orbital and the spin contributions in the $K$-$\chi$ plot. First, within the Korringa relationship, the spin contribution in low-$p$ sample A is negligibly small as compared with those of the other samples. Second, the $K$-$(T_1T)^{-1/2}$ line for each sample, representing the correlation between the $T$-dependences of $K(T)$ and $(T_1T)^{-1/2}$ of a given sample, has a negative slope (dotted lines) for sample C and E and an almost infinite slope for sample A. They do not follow at all the linear behavior with a positive slope expected from the Korringa relationship. This very likely implies the presence of an additional $T$-dependent contribution to $K(T)$ other than $K_{\mathrm{spin}}$, which is small but dominates the slope of $K$-$(T_1T)^{-1/2}$ line for each sample and can be ascribed naturally to $K_{\mathrm{orb}}(T)$.

Let us now return to the $K$-$\chi$ plot in Fig. 4A with the information from Fig. 4B. As $K_{\mathrm{spin}} \sim 0$, the $K$-$\chi$ relationship for sample A should represent that for the non-spin contributions, $K = K_{\mathrm{chem}} + K_{\mathrm{orb}} = K_{\mathrm{chem}} + A_{\mathrm{orb}}\chi_{\mathrm{orb}}$, which is indicated by the red solid line in Fig. 4A and gives an estimate of $A_{\mathrm{orb}} = 88\pm14$ kOe/$\mu_B$. We can ascribe the slopes of the $K$-$\chi$ lines for samples C and E (green and blue dashed lines) close to $A_{\mathrm{orb}} = 88\pm14$ kOe/$\mu_B$ to the predominant orbital contribution, the $T$-dependence of $K_{\mathrm{orb}}(T)$ and $\chi_{\mathrm{orb}}(T)$. The predominance of the orbital contribution $K_{\mathrm{orb}}(T)$ in the $T$-dependence of $K(T)$ can naturally account for the negative slopes of $K$-$(T_1T)^{-1/2}$ for each sample in Fig. 4B. For samples other than sample A, the $T$-dependences of $K_{\mathrm{orb}}(T)$ are weakly negative and those of $(T_1T)^{-1/2}$ are very weakly positive. For sample A, $(T_1T)^{-1/2} \sim 0$ and its $T$-dependence is negligible as compared with the other samples.

The almost parallel and upward shift of the $K$-$\chi$ lines for samples C and E from the non-spin contributions line, $K = K_{\mathrm{chem}} + A_{\mathrm{orb}}\chi_{\mathrm{orb}}$, in Fig. 4A should then represent the superposition of an additional spin contribution $K_{\mathrm{spin}} = A_{\mathrm{spin}}\chi_{\mathrm{spin}}$. As $\chi_{\mathrm{spin}}$ is positive, $A_{\mathrm{spin}}$ should be positive and larger than $A_{\mathrm{orb}}$. The positive sign strongly suggests that $A_{\mathrm{spin}}$ is determined by an $s$ electron-like Fermi contact interaction, which may be induced to the conduction electrons in the Pb $6p$ bands by an $sp$-hybridization or by a strong spin-orbit coupling of Pb (*39*) as discussed in $^{209}$Bi NMR on half-Heusler compounds (*40*). The magnitude of $A_{\mathrm{spin}}$, however, can be determined only with arbitrariness without any further assumption. For any positive $A_{\mathrm{spin}}$ larger than $A_{\mathrm{orb}}$, one can choose $K_{\mathrm{spin}}$ and $\chi_{\mathrm{spin}}$ which satisfy $K_{\mathrm{spin}} = A_{\mathrm{spin}}\chi_{\mathrm{spin}}$ and bring the $K$-$\chi$ lines for samples C and E onto the non-spin contributions line $K = K_{\mathrm{chem}} + A_{\mathrm{orb}}\chi_{\mathrm{orb}}$ (red line) with the shift of $-K_{\mathrm{spin}}$ and $-\chi_{\mathrm{spin}}$ (see also fig. S5A). The shifted $K$-$\chi$ points on the non-spin contributions line represent $K = K_{\mathrm{chem}} + A_{\mathrm{orb}}\chi_{\mathrm{orb}}$ of each sample.

To further narrow down the choice of $A_{\mathrm{spin}}$, $K_{\mathrm{spin}}$ and $\chi_{\mathrm{spin}}$, the assumption of $K_{\mathrm{orb}} = A_{\mathrm{orb}}\chi_{\mathrm{orb}} \sim 0$ for samples C and E may be reasonable, as their Fermi levels are located outside the Dirac bands. Then the $K$-$\chi$ relationship over samples C and E should constitute a universal non-orbital contributions line, $K = K_{\mathrm{chem}} + A_{\mathrm{spin}}\chi_{\mathrm{spin}}$, neglecting the small temperature dependence of $K$ and $\chi$. The crossing point between the non-orbital contributions line and the non-spin line $K = K_{\mathrm{chem}} + A_{\mathrm{orb}}\chi_{\mathrm{orb}}$ (red line in Fig. 4A) corresponds to $K_{\mathrm{chem}}$ and the corresponding offset susceptibility $\chi_0 (= \chi - \chi_{\mathrm{spin}} - \chi_{\mathrm{orb}})$. The non-orbital contributions line $K = K_{\mathrm{chem}} + A_{\mathrm{spin}}\chi_{\mathrm{spin}}$ with the assumption of $K_{\mathrm{orb}} = 0$ for samples C and E can be roughly drawn as the black broken line in Fig. 4A. This non-orbital line and the crossing point with the non-spin line in Fig. 4A yields estimates of $A_{spin}$ = 210 kOe/$\mu_B$, $K_{chem}$ = 0.055% (gray horizontal line), $\chi_0$ = –8.1×10$^{-5}$ emu/mol (gray vertical line), and $K_{spin}$ ~ 0.042% and ~ 0.102% for samples C and E, respectively (blue vertical arrows). As shown in Fig. 4B, the estimated $K_{spin}$ (blue vertical arrows) and $K_{chem}$ (horizontal line) from Fig. 4A are fully consistent with those estimated from the Korringa relation with $S^*$ = 6.3$S$ where $K_{orb}$ = 0 is also assumed. The $\chi_0$ obtained above represents the core contribution to $\chi$ and agrees well with the core susceptibility of –8.5×10$^{-5}$ emu/mol estimated from the atomic values in the literature (*32*), which justifies the assumption of $K_{orb}$ = 0 as the first approximation.

Using the $K_{chem}$ and the $\chi_0$ consistently determined for $K$-$\chi$ and $K$-$(T_1T)^{-1/2}$ in Figs. 4A and 4B, a large diamagnetic orbital contribution in the Knight shift $K$ and the bulk magnetic susceptibility $\chi$, $K_{orb}$ = –0.09% and $\chi_{orb}$ = –5.4×10$^{-5}$ emu/mol at 70 K and $K_{orb}$ = –0.06% and $\chi_{orb}$ = –3.6×10$^{-5}$ emu/mol at 300 K, are estimated for sample A. These orbital contributions apparently dominate the distinct diamagnetism in sample A of which Fermi level lies in the Dirac bands.

The estimated $A_{orb}$ ~ 88 kOe/$\mu_B$ from the $K$-$\chi$ plot implies the unconventional character of giant orbital diamagnetism. $K_{orb}$ is normally driven by a van Vleck paramagnetic susceptibility with $A_{orb}$ = 2$\langle r^{-3} \rangle$ (*41*) determined by the distance $r$ between the nuclei and the orbiting electrons. $A_{orb}$ ~ 2000 kOe/$\mu_B$ is estimated for 6$p$ orbitals of Pb (*42*), which is one order of magnitude larger than the experimentally observed $A_{orb}$ ~ 88 kOe/$\mu_B$ for the Dirac semimetal Sr$_3$PbO. The hybridization of Pb 6$p$ Dirac holes with Sr 4$d$ states and other orbital states could reduce the calculated $A_{orb}$ but not an order of magnitude. The small $A_{orb}$ ~ 88 kOe/$\mu_B$ therefore implies that the orbiting of spatially spread itinerant electrons, not of those completely confined within the atomic orbitals, is in charge of the observed large orbital diamagnetism. If Dirac electrons were uniform free electron gas and not confined in the atomic orbital at all, on the other hand, we would have an estimate of $A_{orb}$ < 1 kOe/$\mu_B$ (*43*), orders of magnitude smaller than the experimentally observed $A_{orb}$ ~ 88 kOe/$\mu_B$. The drastic enhancement from the free electron estimate is reasonable as the Dirac electrons in Sr$_3$PbO are not completely free from the atomic orbital and hopping from one atomic orbital to the others. These comparisons are fully consistent with the theoretical picture of giant orbital diamagnetism based on the inter-band mixing of itinerant Dirac electrons on the crystal lattice, which is distinct from the conventional orbital magnetism of Van Vleck type.

The hole-concentration $p$ (hence Fermi level $E_F$) dependence of the magnetic susceptibility $\chi(T)$ in Fig. 2F and the predominance of the orbital contribution in the enhanced diamagnetism are reproduced by a theoretical calculation of the magnetic susceptibility $\chi^{cal}$ based on the expression in Ref. (9), which explicitly includes the interband effects. Figure 5 indicates $\chi^{cal}$ (solid lines) and its deconvoluted orbital component $\chi^{cal}_{orb}$ (broken lines) as a function of $E_F$ at $T = 232$ K and 348 K, calculated for the tight binding bands of $Sr_3PbO$ (27). Note that the calculated $\chi^{cal}$ does not include the contribution from the core electrons $\chi_0$, a $p$- and $T$-independent constant, and therefore represents $\chi - \chi_0$. An enhanced diamagnetism in $\chi^{cal}$ shows up when $E_F$ is in the Dirac band. Apparently, the orbital component $\chi^{cal}_{orb}$ dominates the enhanced diamagnetism. When $E_F$ lies below the Dirac band regime, the calculated $\chi^{cal}_{orb}$ is much smaller than that in the Dirac band regime, which justifies the assumption of $K_{orb} \sim 0$ for samples C and E. A small but appreciable $T$-dependence of $\chi^{cal}$ is seen particularly in and near the Dirac band regime in Fig. 5, which originates from the orbital contribution $\chi^{cal}_{orb}(T)$ and changes sign from positive to negative upon going away from the mass gap. This is consistent with the increase and the decrease of experimental $\chi_{orb}(T)$ and $\chi(T)$ with increasing $T$ for sample A ($E_F = -45$ meV) and sample C ($E_F = -125$ meV), respectively. These qualitative agreements between the theory and the experiment provide a further support for the validity of the above analysis of NMR results.

Quantitatively, however, the calculation based on the tight binding model does not allow us to capture the details of experimental results. $\chi^{cal}$ for $E_F = -45$ meV (corresponding to sample A) shows an additional diamagnetic contribution of $\sim -2 \times 10^{-4}$ emu/mol (essentially orbital in origin) as compared to those for $E_F < -100$ meV (samples B-E). This is almost a factor of four larger than the experimental orbital contribution $\chi_{orb} \sim 0.5 \times 10^{-4}$ emu/mol for sample A, which is difficult to account for only by the strong $E_F$-dependence of $\chi^{cal}$ and the ambiguity in the estimate of $E_F$. $\chi^{cal}_{orb}$ and $\chi^{cal}$ appear to be overestimated within the framework of the present calculation.

In conclusion, our $^{207}$Pb NMR study of the 3D Dirac electron system $Sr_3PbO$ antiperovskite clearly revealed the orbital origin of large diamagnetism observed in the bulk magnetic susceptibility when its $E_F$ lies in the Dirac bands. This orbital diamagnetism is distinct from the ordinary orbital magnetism in that the orbiting electrons are not confined within the atomic orbitals but hop between the atomic orbitals. These observations are fully consistent with the microscopic picture of giant orbital diamagnetism of Dirac electrons established theoretically after the debates over decades and provide the first firm experimental evidence of such. The calculated orbital susceptibilities as a function of $E_\text{F}$ and $T$, based on the theories, indeed reproduce qualitatively the experimentally isolated orbital contribution to the magnetic susceptibility. Our results open up a fascinating possibility to further explore not only the intra-band effects but also the inter-band effects in topological semimetals.

# Figures

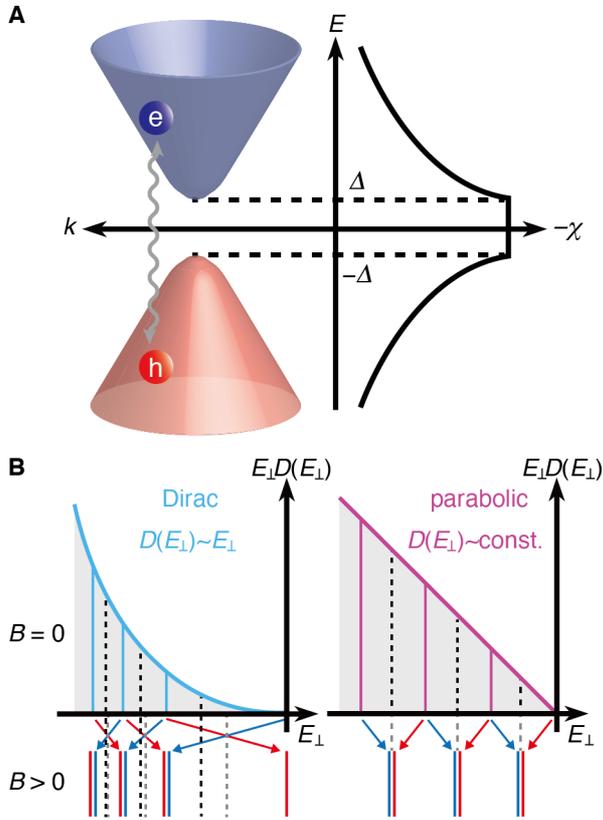

**Fig. 1 Giant orbital diamagnetism of Dirac electrons.** (**A**) Schematic energy dispersion of a massive Dirac electron band (left) and the expected giant orbital diamagnetism as a function of the Fermi level $E_F$ (right). $\varDelta$ is a Dirac mass gap. The orbital diamagnetism takes a maximum with $E_F$ in the Dirac mass gap. (**B**) Energy density $E_\perp D(E_\perp)$ as a function of $E_\perp$ for a Dirac band (left) and an ordinary parabolic band (right). $E_\perp$ and $D(E_\perp)$ represent the kinetic energy originating from the two-dimensional momentum perpendicular to the applied field and the density of states as a function of $E_\perp$, respectively. The Landau levels in a magnetic field with spin up and down are indicated by the blue and the red lines respectively. The average $E_\perp$ of the Landau levels in a magnetic field, to which the gray shaded area of the zero field $D(E_\perp)$ condenses, is indicated by the gray broken lines. They are larger than the average $E_\perp$ of the corresponding gray shaded area (black broken line) for the Dirac band but identical to that for the normal parabolic band.

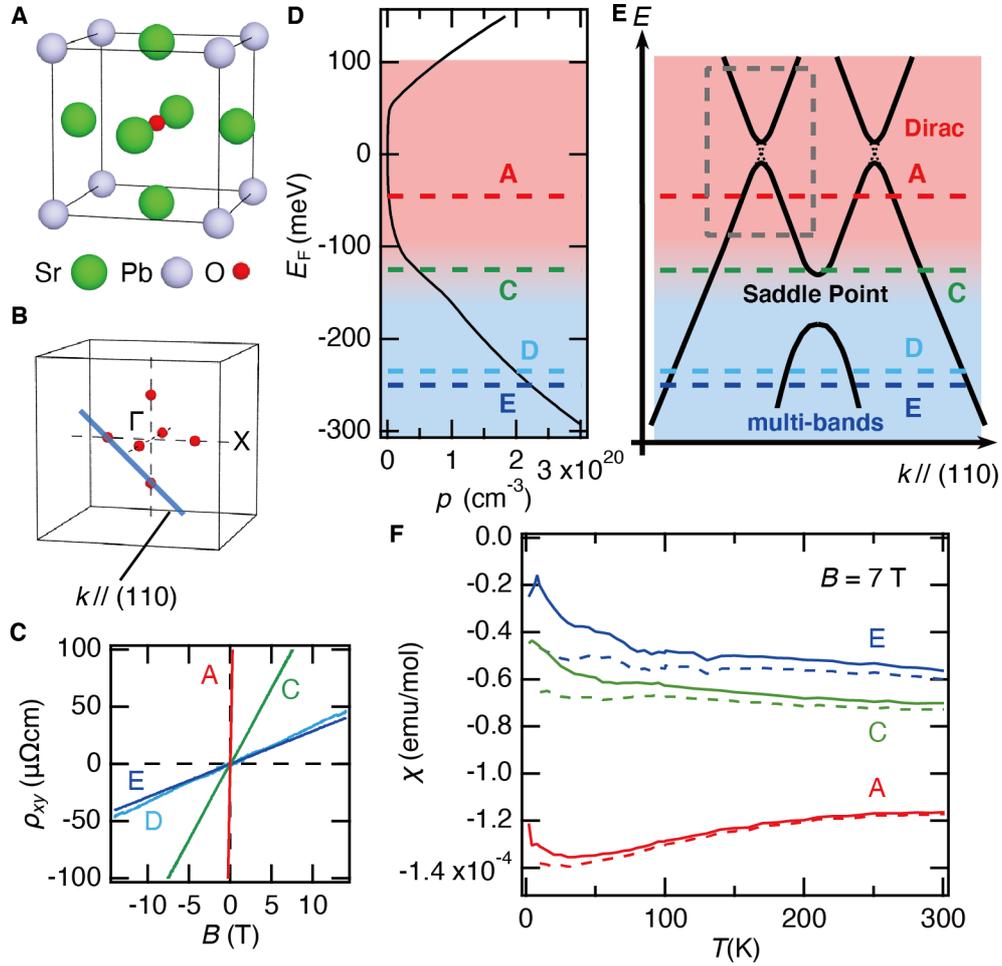

**Fig. 2 Basic electronic structure and bulk magnetic susceptibility in Sr$_3$PbO antiperovskite.** (**A** and **B**) Crystal structure and the first Brillouin zone for Sr$_3$PbO antiperovskite. Six Dirac points (red points in (B)) are protected by $C_4$ rotational symmetry along Γ-X lines. The blue line parallel to the (110) direction connects two Dirac points, between which a saddle point (SP) exists. (**C**) Field dependence of Hall resistivity $\rho_{xy}$ in the zero-field limit. The slopes yield the hole densities $1.6 \times 10^{18}$, $5.0 \times 10^{19}$, $2.0 \times 10^{20}$ and $2.2 \times 10^{20}$ cm$^{-3}$ for samples A, C, D and E, respectively. (**D**) Total carrier density as a function of the Fermi energy ($E_F$) obtained from a band calculation, which gives the estimates of $E_F$ (dashed lines) for samples A, C, D and E as –45, –125, –235 and –250 meV respectively from the experimental hole densities obtained from (C). (**E**) Schematic band structure of Sr$_3$PbO antiperovskite for a $k$-axis along the blue line in (B), which is divided into three regimes, the Dirac bands (red area), the saddle point (SP) and the multi-bands (blue area). The Dirac bands enclosed by the gray dashed rectangle give rise to the giant orbital diamagnetism as illustrated in Fig. 1A. (**F**) Temperature dependence of magnetic susceptibility $\chi$ for samples A, C and E (solid lines). The magnitude of diamagnetic susceptibility increases with decreasing the hole density $p$, in particular very rapidly from sample C to A. By subtracting Curie-like contributions at low temperatures, the intrinsic behaviors of $\chi$ are estimated (broken lines).

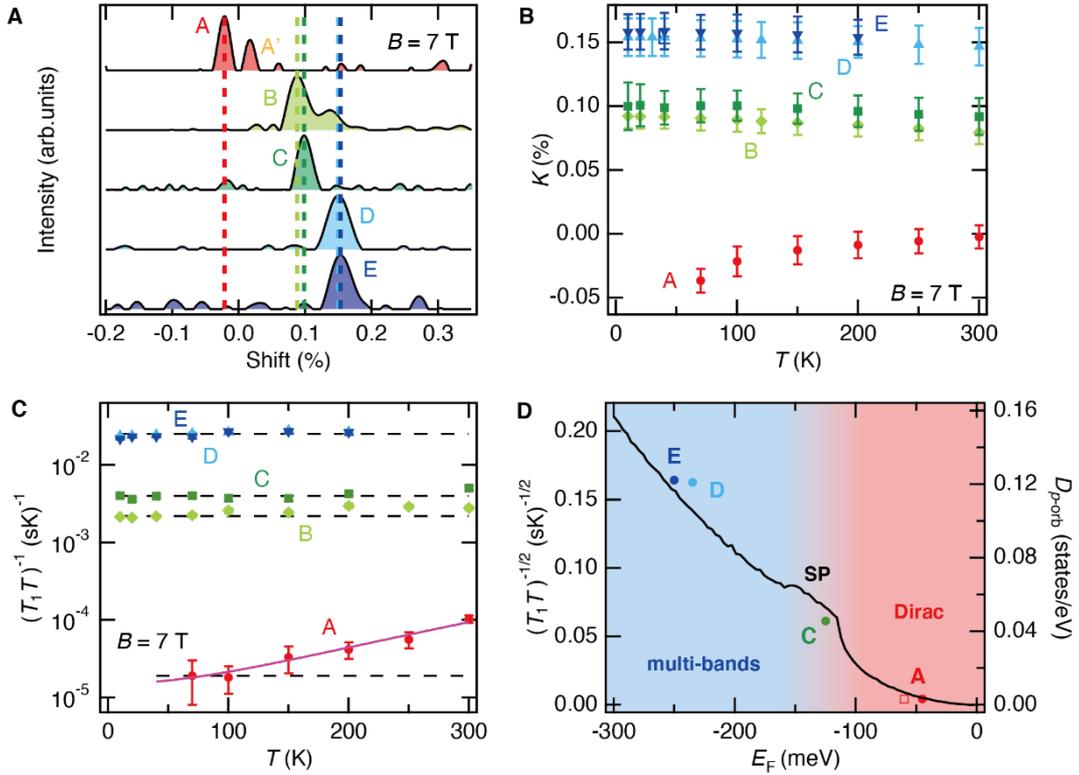

**Fig. 3 NMR spectra, Knight shift $K$ and spin lattice relaxation rate $T_1^{-1}$ in $Sr_3PbO$ antiperovskite.** (**A**) NMR spectra for samples A-E at a temperature $T = 150$ K. The peak positions systematically shift to the negative side upon decreasing the hole density $p$. The hole density $p$ and the Fermi level $E_F$ for each sample are displayed in Figs. 2D and 2E. (**B**) Temperature dependence of the NMR Knight shift $K(T)$ for samples A-E. Note the positive correlation between $K(T)$ and $\chi(T)$ in Fig. 2F is indicative of the positive hyperfine coupling constants $A_{spin}$ and $A_{orb}$. (**C**) Temperature dependencies of the spin-lattice relaxation rate divided by temperature $(T_1T)^{-1}(T)$ for samples A-E. The Korringa law, $(T_1T)^{-1}$ = constant (black dashed line) holds well. A crossover from $T$-independence to $T^2$- dependence for sample A reflects the strongly energy dependent density of states of Dirac electrons. (**D**) Fermi energy $E_F$-dependence of $(T_1T)^{-1/2}$ at 100 K (closed circles), which is scaled well with the calculated partial density of states for Pb $6p$-orbitals at $E_F$, $D_{p\text{-orb}}$ (black line).

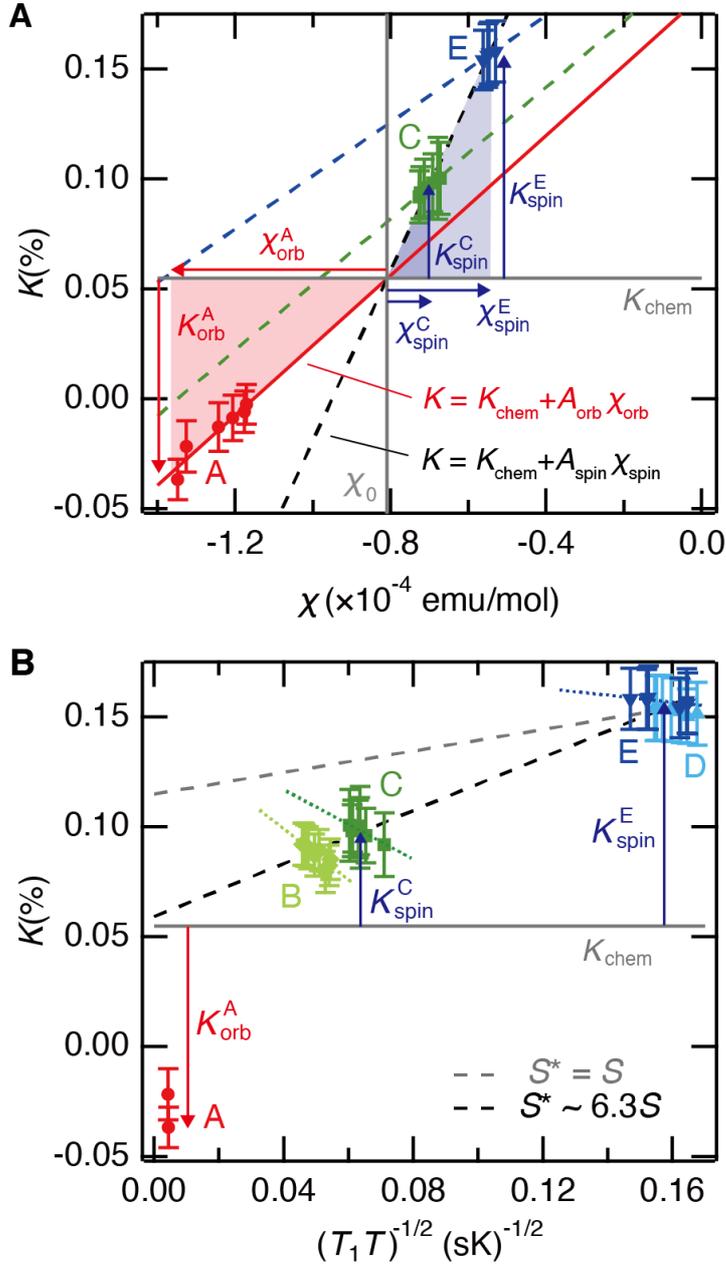

**Fig. 4 Separation of Knight shift $K(T)$ into the spin and the orbital contributions.** (A) Knight shift $K(T)$ vs magnetic susceptibility $\chi(T)$ plot for samples A, C and E. The hole density $p$ and the Fermi level $E_F$ for each sample are displayed in Figs. 2D and 2E. Here the intrinsic $\chi(T)$ after the subtraction of the Curie contribution (broken line in Fig. 2F) was used for the plot. The orbital hyperfine coupling constant $A_{orb} = 88\pm14$ kOe/$\mu_B$ can be estimated from a linear fit to the $K(T)$-$\chi(T)$ relationship for sample A (red line), where the spin contribution $K_{spin}$ is almost zero. An upward deviation for samples C and E from the red line can be attributed to $K_{spin}$. The black broken line indicates the estimated $K_{spin}$ with the assumption of $K_{orb} \sim 0$. The crossing point between the red line and the black

broken line represents the chemical shift $K_{chem}$ and the core magnetic susceptibility $\chi_0$. (See the main text.) (**B**) Knight shift $K(T)$ as a function of the spin-lattice relaxation rate $(T_1T)^{-1/2}$ for sample E to sample A. The spin contribution $K_{spin}$ expected from the Korringa relationship $T_1TK_{spin}^2 = S$ is indicated by the gray broken line. The black broken line indicates a modified Korringa relationship with an enhanced Korringa constant $S^* \sim 6.3S$, which assumes a dominant spin contribution and hence almost zero orbital contribution, $K_{orb} \sim 0$, for heavily doped samples B-E. Note that the extrapolation of the black dashed line to $(T_1T)^{-1/2} = 0$ gives an estimate of the chemical shift $K_{chem}$ with the assumption of $K_{orb} \sim 0$ for samples B-E. The large and additional negative shift of sample A should be ascribed to the orbital contribution, $K_{orb}$.

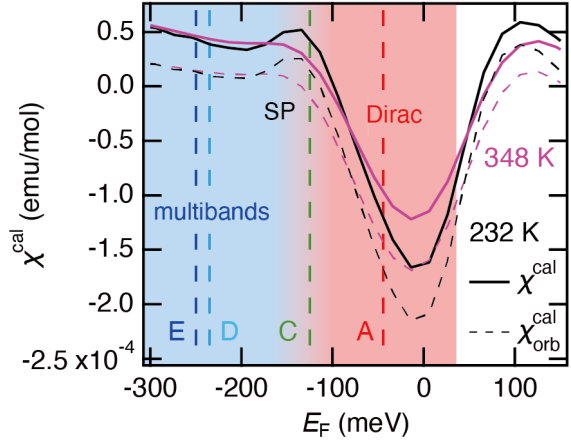

**Fig. 5 Theoretical calculation of magnetic susceptibility as a function of the Fermi level $E_F$ for $Sr_3PbO$ antiperovskite.** The Fermi level $E_F$-dependence of magnetic susceptibility $\chi^{cal}$ at 232 K and 348 K (solid lines), calculated for tight-binding bands of $Sr_3PbO$ by incorporating the inter-band effect. The diamagnetic contributions from the core electrons $\chi_0$, which are temperature and hole density independent constants, are not included in the calculation. To compare with the experimental results, $\chi^{cal}+\chi_0$ should be used. A large diamagnetism grows with approaching the Dirac mass gap. The deconvoluted orbital contribution $\chi^{cal}_{orb}$ at 232 K and 348 K are shown by the broken lines. The experimentally determined $E_F$ for samples A-E are indicated by the vertical broken lines.

**Acknowledgements**

We thank Dennis Hwang, Oleg Sushkov, T. Hirosawa and H. Maebashi for fruitful discussions and comments, and K. Pflaum for technical assistance. This work was supported by Japan Society for the Promotion of Science (JSPS) KAKENHI (Grants No. 24224010, No. 15K13523, No. JP15H05852, No. JP15K21717 and No. 17H01140) and Alexander von Humboldt foundation. S.S. acknowledges financial support by JSPS and the Materials Education program for the future leaders in Research, Industry, and Technology (MERIT).

**Author contribution**

J.N and C.M. synthesized single crystals. S.S. and A.W.R. characterized the crystals. S.S., K.K. and A.W.R. performed the magnetic susceptibility measurements and S.S. and K.K carried out the NMR experiments. T.K. and M.O. calculated the band structure